\documentclass[useAMS,usegraphicx,usenatbib]{mn2e}
\usepackage{graphicx,deluxetable}
\newcommand{\apj}{ApJ}
\newcommand{\aj}{AJ}
\newcommand{\apjl}{ApJL}
\newcommand{\mnras}{MNRAS}
\newcommand{\apjs}{ApJS}
\newcommand{\araa}{ARA\&A}
\newcommand{\aap}{A\&A}

\newcommand{\nat}{Nature}

\newcommand{\pasp}{PASP}
\newcommand{\ssr}{SSR}
\newcommand{\jcap}{JCAP}
\begin{document}
\title[Structure and Turbulence in Simulated Galaxy Clusters]{Structure and Turbulence in Simulated Galaxy Clusters and the
Implications for the Formation of Radio Halos}
\author[E.J. Hallman and T.E. Jeltema]{Eric
  J. Hallman$^{1}$\thanks{E-mail: ehallman@cfa.harvard.edu} and Tesla E. Jeltema$^{2}$\\
$^{1}$Harvard-Smithsonian Center for Astrophysics, Cambridge, MA 02138\\
$^{2}$Santa Cruz Institute for Particle Physics and UCO/Lick
Observatories, 1156 High St., Santa Cruz, CA 95064}

\maketitle
\label{firstpage}
\begin{abstract}
We track the histories of massive clusters of galaxies formed within a
cosmological hydrodynamic simulation.  Specifically, we track the time
evolution of the energy in random bulk motions of the intracluster
medium and X-ray measures of cluster structure and their relationship
to cluster mergers.  We aim to assess the viability of the turbulent
re-acceleration model for the generation of giant radio halos by
comparing the level of turbulent kinetic energy in simulated clusters
with the observed properties of radio halo clusters, giving particular
attention to the association of radio halos to clusters with disturbed
X-ray structures.  The evolution of X-ray cluster structure and
turbulence kinetic energy, $k$, in simulations can then inform us
about the expected lifetime of radio halos and the fraction of
clusters as a function of redshift expected to host them.  We find
strong statistical correlation of disturbed structure measures and the
presence of enhancements in $k$.  Specifically, quantitatively
``disturbed'', radio halo-like X-ray morphology in our sample
indicates a 92\% chance of the cluster in question having $k$ elevated
to more than twice its minimum value over the cluster's life.  The
typical lifetime of episodes of elevated turbulence is on the order of
1 Gyr, though these periods can last 5 Gyrs or more.  This variation
reflects the wide range of cluster histories; while some clusters
undergo complex and repeated mergers spending a majority of their time
in elevated $k$ states, other clusters are relaxed over nearly their
entire history.  We do not find a bimodal relationship between cluster
X-ray luminosity and the total energy in turbulence that might account
directly for a bimodal $L_X-P_{1.4 GHz}$ relation. However, our result may be
consistent with the observed bimodality, as here we are not including a full treatment of cosmic rays sources and magnetic fields.  

\end{abstract}

\section{INTRODUCTION}
Galaxy clusters have long been known as radio synchrotron sources.
Given that observations show that the intracluster medium is
typically magnetized at the $\sim$$\mu G$ level
\citep[see][]{carilli,govoni_feretti}, what is required for the radio emission is a nonthermal population of
accelerated electrons.  Arguments for various scenarios responsible
for cluster radio emission therefore hinge on the method of particle
acceleration, using the observed radio and X-ray properties of galaxy
clusters as discriminatory evidence. 
Radio properties of the sources in galaxy clusters vary, though in
general the types of sources can be separated into two basic
categories, sources associated with individual galaxies and those
associated with the diffuse intracluster medium (ICM).  Those
associated to individual galaxies include the jet and lobe sources
connected to active galactic nuclei (AGN) \citep[e.g.,][]{hardcastle,nulsen}. The diffuse sources can
be further divided into two broad classes, radio relics and radio
halos.  Radio relics include both steep spectrum re-energized AGN jet
lobes, as well as polarized sources likely associated with recently
accelerated cosmic-ray (CR) electrons at merger shocks \citep[e.g.,][]{clarke_enss,vanweeren}. These
extended, elongated features are steep spectrum sources with curved shapes, spatially distinct and
strongly argued as coincident with shocks from both observations and
numerical simulations \citep{roett3667,kassim01,bagchi,skillman10}.  Radio halos (RHs), by contrast, are Mpc-scale,
diffuse, unpolarized, steep spectrum radio emitting regions typically centered on
the X-ray center of the ICM, with a spatial distribution similar to
the X-ray emission \citep{liang,feretti,giacintucci,macario}.  

The origin of RHs is
less certain than the radio relics, though it was recognized a decade
ago \citep[e.g.][]{buote}, and observational evidence now continues to suggest
that they are associated with merging clusters \citep{sarazin04,govoni04,brunetti09,cassano}.  For example, the recent work
of \citet{cassano} uses radio observations (with GMRT) and
\textit{Chandra} observations of an X-ray selected sample of galaxy
clusters to make a strong case for a division in the observed X-ray
morphological properties of RH and non-RH galaxy clusters. In brief,
clusters that are more disturbed in the X-ray power ratios,
centroid shift measures and have lower concentration parameters are those that tend to host radio halos.
Those with relaxed structure measures and higher concentrations tend to not have detectable
large-scale radio emission.

Recent
observational and simulation work argues that RHs are created through merger-induced turbulent
re-acceleration of an existing population of CRs in the ICM. The argument for turbulent re-acceleration of a pre-existing CR
electron population as the cause of RHs appeals to a combination of
observational properties. First, the radio emission is broadly
distributed over a relatively large volume, roughly centered on the
X-ray centroid of the cluster, and more or less spatially uniform
\citep[for a recent review, see][]{ferrari}.
Given the short synchrotron lifetimes of the CR electrons responsible
for the emission, they must be accelerated \textit{in situ}, ruling
out acceleration from large individual shock structures on
morphological grounds \citep{brunetti01}.  Turbulence provides a source of energy that is
distributed relatively evenly over the cluster volume.  Second, this model naturally explains the connection of the presence of RHs to cluster mergers.  Though second
order Fermi processes are usually invoked \citep{petrosian}, there
are legitimate questions about the exact method by which the turbulent
energy can be converted into accelerated particles.  In addition, a
key prediction of the turbulent re-acceleration model is the expected
presence of very steep spectrum RHs ($\alpha > 1.6$)
\citep{cassano_proc}.  Two recent RHs have been found to host so
called ultra steep spectrum radio halos, A521 \citep{brunetti08} and
A697 \citep{macario}.  These discoveries, though small in number so
far, fit the turbulent re-acceleration model. 

Another explanation proposed for the origin of RHs is hadronic production. This model invokes the production of secondary relativistic electrons generated by the
collisions of relativistic protons and the thermal population of
atomic nuclei in the ICM.  In the hadronic scenario, the secondary electrons
are responsible for the observed radio emission \citep[e.g.,][]{pfrom08,keshet_loeb,keshet_secondary},
and the proponents argue that the hadronic model more naturally
explains the correlation of the radio emission with the X-ray emitting ICM (i.e. the target population for p-p collisions) and the presence of both RHs and radio relics in clusters. Merger induced enhancement of cluster magnetic fields is invoked to explain the RH-merger connection \citep{kushnir,keshet_loeb}. However, recent $\gamma$-ray upper limits from \textit{Fermi}
observations of galaxy clusters are beginning to constrain the
capability of hadronic models to account for RHs
\citep[see][]{jeltema_profumo}, as the hadronic processes should also
generate $\gamma$-rays from neutral pion decay. The recent discovery
of ultra steep spectrum radio sources described above is also not
predicted in the hadronic scenario given the energetic considerations
required by the $\gamma$-ray upper limits. These and other
constraints suggested by observations of RHs, such as the spectral and
spatial properties of the radio emission \citep[see][]{brunetti04}, argue
against the hadronic model. 

One more piece of observational evidence in the case of radio halos is the
relationship between the bulk X-ray and radio properties of massive galaxy
clusters. The radio luminosity at 1.4$GHz$ and the X-ray luminosity of
massive radio halo galaxy clusters exhibit a strong correlation for many
objects.  However, there apparently exists a bimodal distribution in the $L_x -
P_{1.4GHz}$ plane, where many clusters have no radio detections,
only upper limits in $P_{1.4GHz}$ that lie well below the scaling
relation. These objects are of similar mass and X-ray luminosity to
those that clearly host radio halos and are on the scaling relation.  There is a
significant gap between the measured radio luminosity of the RHs and
the upper limits of the non-RH objects, which suggests that the
transition between the two phases (radio halo versus non-RH) must be
relatively quick \citep{brunetti09}.  While there were initial
concerns that the bimodality of the $L_x -
P_{1.4GHz}$  relation might be an artifact of selection effects
\citep[e.g.,][]{rudnick}, the depth of the radio data for the current
samples appears to address that issue. The presence of a strong
correlation of radio power with X-ray luminosity for RH clusters, the
bimodality of the relation, and the inferred lifetime and transition
times for RHs have been used to argue for or against models of radio
halo production \citep[e.g.][]{kushnir,brunetti09}.  In this case,
numerical simulations provide a window into the evolution of
the relevant physical properties of the ICM. 

In this paper, we study the evolution over time of massive clusters formed within a cosmological volume simulation, which allows us to track cluster mergers, the evolution of cluster structure, and the presence and evolution of turbulence.  Using simulations, we can investigate the frequency and timescale over which clusters exhibit elevated levels of turbulence and X-ray structures similar to those observed for RH clusters. The use of cosmological simulations turns out to be critical to accurately capture cluster merger histories.
%The main idea is that clusters with recent
%mergers exhibit these X-ray properties, and that the turbulence
%generated in the merger is the source of free energy for the
%acceleration of CR electrons, and is responsible for the radio halo. However,
%given the size of their sample ($n$=32) and the lack of information about
%the temporal evolution of the cluster properties, it is difficult to
%make broad generalizations about the details of the process. In this work, 
In particular, we attempt to understand the connection between the X-ray
morphology of galaxy clusters and the putative energy source for CR
acceleration in the ICM: turbulence.  We do not attempt to determine the method of particle acceleration, we merely discuss the probability of the turbulent
re-acceleration model of RHs given the correlations between the physical properties
of the simulated clusters and their observed morphological
characteristics.  We study both the ``snapshot''
correlations for our clusters between the X-ray structure measures
(power ratios and centroid shifts) and
the turbulence kinetic energy and the time evolution of these
properties. 

Section 2 of this paper describes the methodology of the analysis we
undertake, including the details of the numerical simulations and
analysis and the method of calculating the X-ray structure measures
from the synthetic observations. Section 3 describes the distribution
of simulated galaxy cluster structures compared to observations, the time evolution of both the structure
measures and turbulence kinetic energy, and compares how the turbulence kinetic energy correlates with the observed properties of radio halo clusters.  Section 4 gives estimates for
lifetimes and duty cycles for radio halos based on the typical duration of episodes of elevated turbulence kinetic energy and the fraction of clusters in these states.  Section 5 summarizes and
discusses the implications of our results. 
\section{METHODOLOGY}

\subsection{Simulations}
For this study we use a numerical cosmological hydrodynamic/N-body
simulation generated using the \texttt{Enzo}\footnote{http://enzo.googlecode.com/} code, a publicly
available adaptive mesh refinement (AMR)
cosmology code developed by Greg Bryan and colleagues \citep{bryan97,bryan99,norman99,oshea04,
2005ApJS..160....1O}.
The specifics of the Enzo code are described in detail in these papers
(and references therein).

 This simulation is set up as follows.  We initialize
our calculation at $z=99$ assuming a cosmological model with  $\Omega_m = 0.27$, 
$\Omega_b = 0.04$, $\Omega_\Lambda = 0.73$, $h=0.7$ (in units of 100 km/s/Mpc), 
$\sigma_8 = 0.9$, and using an \citet{eishu99} power spectrum
with a spectral index of $n_s = 0.97$.  The simulation is of a volume of the 
universe 128~h$^{-1}$~Mpc (comoving) on a side with a $256^3$ root grid.  The dark matter
particle mass is $9.1 \times 10^{9}$~h$^{-1}$~M$_\odot$.  The simulation was then evolved to $z=0$ with
a maximum of $5$ levels of adaptive mesh refinement (a maximum spatial resolution of 
$15.6$~h$^{-1}$ comoving kpc), refining on dark matter 
and baryon overdensities of $8.0$.

This simulation includes a prescription for radiative cooling of the gas
using non-equilibrium cooling and chemistry for H and He, and a spatially uniform but time-varying Haardt-Madau
ultraviolet radiation background.  In this case, we have not included
metal line cooling as has been done in previous calculations, since in
this study we are interested in the morphological properties of the
clusters.  Many investigators have documented the over-cooling
problems associated with including metal cooling in cosmological
simulations, and in previous work we have shown that this over-cooling
leads to a severe mismatch of structure measures between the observed
and simulated cluster images. 

We perform all our post-processing analysis for this work using the \texttt{YT}\footnote{http://yt.enzotools.org/}  analysis
toolkit \citep{turk_yt}. For each of 132 simulation outputs equally
spaced in time ($\delta t$ = 0.055 Gyr) in the
redshift interval 0 $\leq z \leq  $0.9,  we ran the HOP halo-finding
algorithm \citep{eishut98} on the 
dark matter particle distribution to produce a dark matter halo
catalog. For each halo catalog in each time interval, we create
spherically averaged radial profiles of a set of physical properties
(including density and mass) to calculate the value of $r_{200}$ and
$M_{200}$ for the identified halos.  $M_{200}$ refers to the total mass inside
a radius of $r_{200}$, the radius at which the overdensity average
inside the sphere centered on the cluster is 200 times the critical
density. We then took all the halos at $z$=0 with $M_{200} \geq 3 \times 10^{14}
M_{\odot}$ (a relatively arbitrary cutoff) as our sample. We choose
the high mass objects so that we can make a reasonable comparison to
the very high mass objects of \citet{cassano}, and to limit our sample
to the best resolved objects in the simulation. This cut results in a
sample at $z$=0 of 16 galaxy clusters.  

Following the identification of the sample, we generated a set of
synthetic 0.3-8.0 keV X-ray images of each cluster generated from the
\texttt{Cloudy}\footnote{http://nublado.org/} software \citep{ferland}. We then make a corresponding image of each of these
clusters from each of the 132 time outputs described above.  Each
initial image represents a projected area of 8 by 8 Mpc around each
cluster in order to study both the cluster and the surrounding
structures. The X-ray images are output in FITS format with the
correct angular scale for their redshift, and with the flux modified
by the distance. Therefore, we are left with a series of synthetic
X-ray surface brightness images that we can analyze in a similar way
to observed images of galaxy clusters. Each series represents a time
history of each of our 16 clusters. Additionally, we have calculated
the bulk properties of each cluster and its most massive predecessors
throughout this time history.

\subsection{Structure Measures and Turbulence}

\subsubsection{Structure Measures}
The association of radio halos with clusters undergoing mergers, known visually for quite some time, has been established quantitatively through the presence of significant substructure or disturbed cluster structure in the X-ray surface brightness distribution \citep[e.g.][]{buote, cassano}. In particular, \citet{cassano} use three structure measures based on X-ray images of the clusters in their sample: the centroid shift $w$, the third order power ratio $P_3/P_0$, and the concentration $c$.  \citet{cassano} show that giant radio halo clusters can be efficiently separated from non-radio halo clusters and radio mini-halos in the plane defined by any two of these structure measures (i.e. $w$ vs. $P_3/P_0$, $c$ vs. $w$, or $c$ vs. $P_3/P_0$).

Here we will employ the former two statistics, the centroid shift and $P_3/P_0$, the calculations of which have been discussed in detail in previous papers \citep{jeltema05, jeltema, hallman_cf}.  In brief, the centroid shift measures the variation (here standard deviation) of the position of the centroid of the X-ray surface brightness of a cluster relative to the X-ray peak in apertures of increasing radius, and w is normalized relative to the largest aperture considered.  The power ratios are based on the multipole moments of the X-ray surface brightness in an aperture of a given size; specifically they are proportional to the sum of the squares of the moments divided by the total cluster flux \citep{bt95}. $P_3/P_0$ is the third order power ratio and is sensitive to deviations from mirror symmetry.  For consistency with \citet{cassano}, we employ the same overall aperture size of 500 kpc for both $w$ and $P_3/P_0$ as well as the same aperture spacing of 25 kpc for the centroid shift (see for example equations 1-5 in \citet{cassano}). In the case of the centroid shift, we remove the central 50 kpc surrounding the X-ray peak from consideration. 
% It turns out there is very little difference with or without the peak.  It must have been an earlier simulation where I was getting problems when including the peak.  
The centroid shift $w$ and the power ratios within an aperture of 500 kpc are calculated for every time output over the history of our 16 simulated clusters.

We do not use the concentration parameter, which is defined by \citet{cassano} as the ratio of the X-ray surface brightness within the central 100 kpc to the surface brightness within 500 kpc.  With this definition, the typical cluster concentration evolves significantly with redshift as a fixed 100 kpc aperture encloses more or less of the cluster volume. This evolution makes a fixed cut for relaxed or disturbed clusters impossible.  A different definition of concentration based on overdensity radii would provide a better criterion for clusters spanning a large redshift range but would not allow a direct comparison to the observed radio halo sample of \citet{cassano}.  We note, however, that in the redshift range spanned by the \citet{cassano} sample, we find a similar median concentration for our simulated clusters to that of the observed clusters once the X-ray peak is removed (as was done for the centroid shifts).

\subsubsection{Turbulence Kinetic Energy}
To evaluate the connection between the morphological structure
measures and the physical state of the cluster, we have made a
calculation of the mass-averaged turbulence kinetic energy (TKE) for each cluster inside
the same radius (500 kpc) that we have calculated the structure
measures. Here we use a common definition of the turbulence kinetic
energy density \citep[see e.g.,][]{choud}
\begin{equation}
k = 0.5( \overline{u_x^2} + \overline{u_y^2} + \overline{u_z^2}),
\end{equation}
where $u_x, u_y, u_z$ are the orthogonal components of the peculiar velocity of the gas in each grid zone
(subtracting the bulk halo velocity), and $k$ is the specific kinetic
energy. The bulk halo velocity is calculated as the mass-weighted
average velocity of all dark matter particles identified by the halo finding
algorithm as halo members. For our analysis we calculate a mass-averaged value of $k$ in
a sphere with $r$=500 kpc around each cluster center.  We note that we are not truly calculating the energy of the full turbulent
cascade.  We can calculate a physical Reynolds number for
flows of characteristic length scale equal to the minimum grid scale (for
grid elements at the peak spatial resolution) of the simulation. Our
grid size at peak spatial resolution and the characteristic parameters of the ICM
flows can be combined into a dimensionless Reynolds number
\begin{equation}
Re = vL/\nu,
\end{equation}
where $v$ is the fluid velocity, $L$ is a characteristic length, and
$\nu$ is the kinematic viscosity of the fluid.  We can estimate the
values for $v$ and $L$ to be roughly the sound speed $c_s$ and minimum
grid scale. Estimates of the viscosity are trickier, but the viscosity for a fully ionized unmagnetized
thermal plasma \citep{braginskii,spitzer} can be written as
\begin{equation}
\mu_{um} = 2.2\times10^{-15} \frac{T^{5/2}}{ln\Lambda} g cm^{-1} s^{-1}.
\end{equation}
For typical values of the ICM temperature ($kT \approx$10 keV) and density and assuming a sound speed of $c_s$=1000 km/s, the Reynolds
number at our peak grid resolution is roughly $Re \approx 250$. 
Flows with these Reynolds numbers are below the expected threshold for
transition from laminar to turbulent flow in an ideal gas.
Additionally, for our
simulations, the details of turbulent flow (should it exist) are not captured at
our smallest grid scale or below.  However, $Re$ scales linearly with the
characteristic length of the flow, and large scale
eddies should be captured.  We expect that turbulence injection in the
ICM happens at scales well above our grid size ($L \approx 100-400
kpc$) \citep{brun_laz}. What we do capture in this calculation is the kinetic energy
associated with bulk random motions of the ICM gas at and above the
grid scale of the simulation.  This motion
would be converted to thermal energy through the turbulent cascade
(and through shock dissipation) in the real ICM. So this calculation
is made to simply characterize the available energy in random bulk
motion of the ICM, which we assume will eventually be dissipated into
heat (and possibly CRs) through the turbulent cascade.  Indeed, recent
work by \citet{paul} indicates that our metric $k$ should be
relatively insensitive to the spatial resolution of the simulation grid.

The quantity $k$ is tracked with time in the history of each cluster to
determine the change in the TKE. If the paradigm that cluster radio halos
are powered by turbulent re-acceleration of CRs is correct, increases in
the TKE should lead to an increase in the radio brightness. Therefore,
we can make estimates of the lifetime of radio halos by tracking the
TKE evolution. We also study the correlation of the X-ray structure
measures with the TKE; since the presence of a radio halo is observed to be 
linked to X-ray structure, the TKE should likewise be correlated to cluster structure if indeed turbulence is connected to the production of radio halos.

\section{CONNECTION OF CLUSTER MERGERS, STRUCTURE, AND TURBULENCE}

\subsection{Distribution of Cluster Structures}

\citet{cassano} show that radio halo clusters can be distinguished
from non-radio halo clusters based on their observed X-ray structure.
In particular, in terms of the two structure measures we consider,
\citet{cassano} find that clusters hosting radio halos tend to lie in
the upper-right quadrant of the $w - P_3/P_0$ plane with high centroid
shifts and high $P_3/P_0$.  %We, therefore, investigate our simulated clusters to see what fraction of the time and during what phases of mergers they have radio halo like cluster structures. 
 Below, we compare the distribution of clusters structures for our simulated clusters to observed clusters from the \citet{cassano} sample and show that they are similar.
 
In Figure 1, we show the distribution of structures of our
simulated clusters in the $w - P_3/P_0$ plane for all time outputs with $0.2 \leq z \leq 0.4$, the redshift range spanned by the
\citet{cassano} sample.  \citet{cassano} make cuts for radio
halo versus non-halo clusters based on the median of each structure measure,
which for their sample gives $w = 0.012$ and $P_3/P_0 = 1.2 \times
10^{-7}$.  Most of the radio halo clusters in their sample do in fact
have $w > 0.012$ and $P_3/P_0 > 1.2 \times 10^{-7}$.  For our
simulated sample in a similar redshift range ($0.2 \leq z \leq 0.4$),
we find median values of $w = 0.010$ and $P_3/P_0 = 4.6 \times
10^{-8}$, similar particularly for the centroid shift to the
\citet{cassano} values.  Both the \citet{cassano} structure cuts (red
lines) and the median values in the simulations (black lines) are
shown in Figure 1.  We find that in the $0.2 \leq z \leq 0.4$ redshift
range, $24\%$ of the simulated cluster images have structures above
the \citet{cassano} cuts %for radio halos 
($w > 0.012$ and $P_3/P_0 >
1.2 \times 10^{-7}$) while $36\%$ have structures above the
median values for the simulated sample.  

Overall there is a good match between the structure of
clusters in our simulations and observed clusters in terms of both the typical cluster structure and range of structure measures. %at least in terms of selecting which clusters are likely to host radio halos.  
Comparing Figure 1 to Figure 1 in \citet{cassano} gives the visual impression
that a larger fraction of simulated clusters lie in the off-diagonal
(lower-right and upper-left) regions of the structure plane implying a
less strong correlation of $w$ and $P_3/P_0$.  In Figure 1, a total of
$26-28\%$ (depending on the cuts used) of the simulated cluster images
lie in the two off-diagonal quadrants compared to 5 out of 32 in the
\citet{cassano} sample, not a significant difference given the sample
size. 

The fraction of simulated clusters in the upper-right quadrant of the $w - P_3/P_0$ plane, 
which have X-ray structures similar to those observed for radio halo clusters, is a good match to the
$29 \pm 9\%$ of X-ray luminous clusters found to host radio halos in
the survey of a complete cluster sample with the Giant Meterwave Radio
Telescope (GMRT) \citep{venturi} (from which the \citet{cassano}
sample is drawn) and to the 12 out of 32 clusters hosting giant radio
halos in the \citet{cassano} sample.

If we consider instead the full redshift range of the simulations
($0.0 \leq z \leq 0.9$), we find that clusters have radio halo like
structures $19\%$ ($31\%$) of the time given the \citet{cassano}
(simulated) structure cuts.  As we will show in later sections, some
clusters spend most of their time in disturbed states
while other clusters are relaxed over nearly all of their history.

\begin{figure}
\includegraphics[width=3.5in]{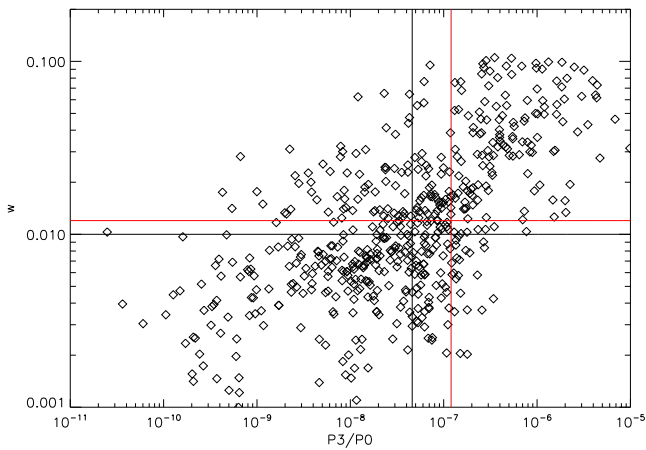}
 \caption{The centroid shift $w$ versus $P_3/P_0$ plane for all the
  simulated cluster images in the redshift range $0.2 \leq z \leq 0.4$
(the range in \citet{cassano}). Red lines indicate the structure cuts from
\citet{cassano}, black are the median values for our simulated
clusters.}
    \label{struc_midz}
\end{figure}

\subsection{Cluster Histories}

%\subsection{Structure and Turbulence in Simulated Clusters}
We follow the evolution of both the structure measures and the TKE for individual simulated clusters through time.  This reveals a wide range of cluster histories.  We investigate the effect of mergers on measured cluster structure (i.e. the observable effects of mergers) and on the TKE.  We can then ask when clusters have structure measures similar to those observed for radio halos and how this relates to the TKE and merger history.

\begin{figure*}
  \includegraphics[width=6.8in]{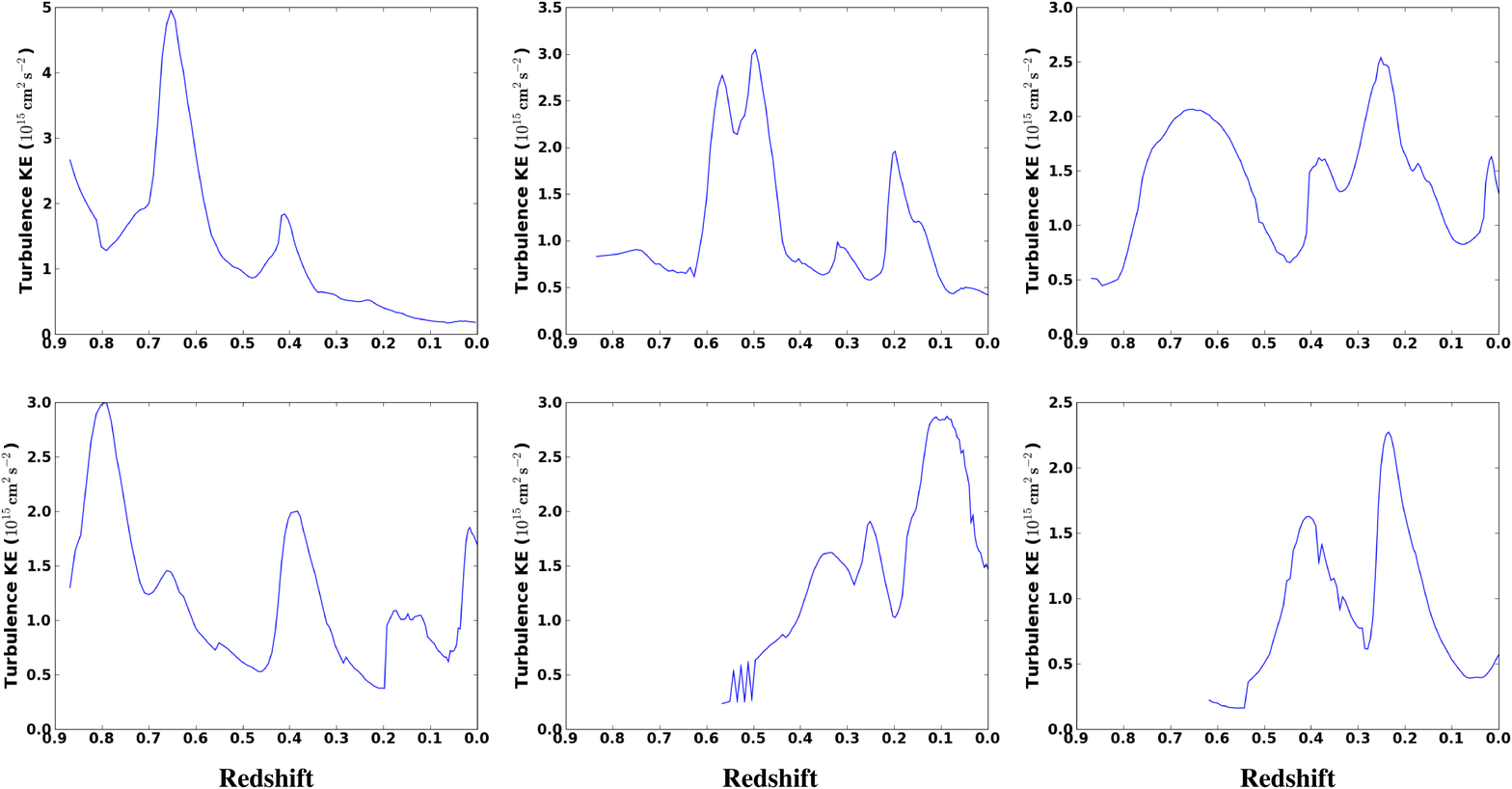}
  \caption{A sample of the turbulence kinetic energy (TKE) density histories
  for 6 of the simulated clusters. These show the wide range of features
  exhibited by the time evolution of the TKE. Note the vertical scale
  differs across the plots.}
\label{all_ke_z}
\end{figure*}

The time histories of the TKE in 6 of the simulated clusters in our sample
are shown in Figure \ref{all_ke_z}.  The clusters are chosen to
represent a wide range of time histories. We use the kinetic energy density
as defined earlier to remove the effect of increasing mass on the
result. In each case, for the purposes of comparison with the
structure measures, we only examine the value of $k$ inside a radius
of 500 kpc.  It is clear from these histories that the value of $k$
undergoes major variations as a function of time in each
cluster. These variations can change the value by factors of 5 or more
over time intervals of 0.5-1 Gyr. The strong increases in the TKE will
be shown in later sections to correspond directly to the incidence of
merging events. These merging events, as has been shown in previous
investigations, also correspond to changes in the X-ray structure
measures. 

%Some basic statistics.  What fraction of the time are clusters in
%disturbed phases in terms of both high structure measures and high
%turbulence?  What fraction of the time do individual clusters spend in
%disturbed phases (i.e. point out that some clusters are almost always
%disturbed while others never are). It might be worth figuring out
%which TKE states are best correlated with the structure plane, but
%from the previous work, it could be hard to tell.  

\begin{figure*}
\begin{center}
\includegraphics[width=6.8in]{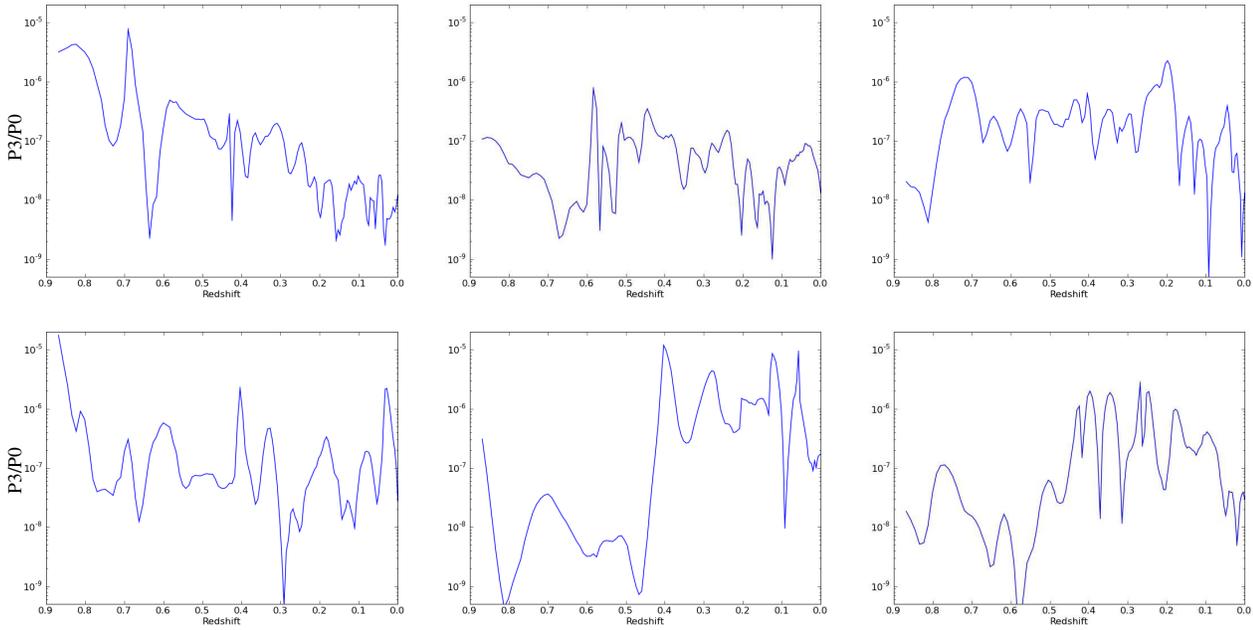}
\end{center}
\caption{A sample of the X-ray $P_3/P_0$ histories
  for the same 6 simulated clusters for which we plotted TKE histories
  in Figure \ref{all_ke_z}. Note the wide variation of $P_3/P_0$ on
  short time scales compared to the smooth variation of the TKE. }
\label{p3_z}
\end{figure*}

However, for the structure measures, the variations with time for
individual clusters are not as smooth as those for the TKE. Figure
\ref{p3_z} shows the variation in the $P_3/P_0$ power ratio as a
function of redshift for the same 6 simulated clusters shown in Figure
\ref{all_ke_z}. While the TKE rises and falls smoothly during merger
events, the value of the X-ray $P_3/P_0$ power ratio jumps up and down
significantly during the whole life of the cluster.  From previous
studies, we were aware that the power ratios can vary significantly on
short time scales.  Merger phase can have a strong effect on the structure
measures.  Strong dips in $P_3/P_0$ are often seen, for
example, at the moment of core passage in major mergers. The X-ray structure measures tend to be highest in the pre-merger and post-core passage re-expansion phases when subclusters are well-separated and/or asymmetric features are most prominent.  Statistically, X-ray structure measures like $w$ and $P_3/P_0$ correlate strongly with cluster dynamical state, but significant scatter is present for individual cluster measurements (projection effects also contribute to the scatter) \citep{jeltema}. Therefore, though a correlation between the turbulence kinetic energy and the
X-ray structure measures is expected, it is not immediately obvious from
individual cluster time histories that this is the case. 

\begin{figure*}
\begin{center}
\includegraphics[width=5.0in]{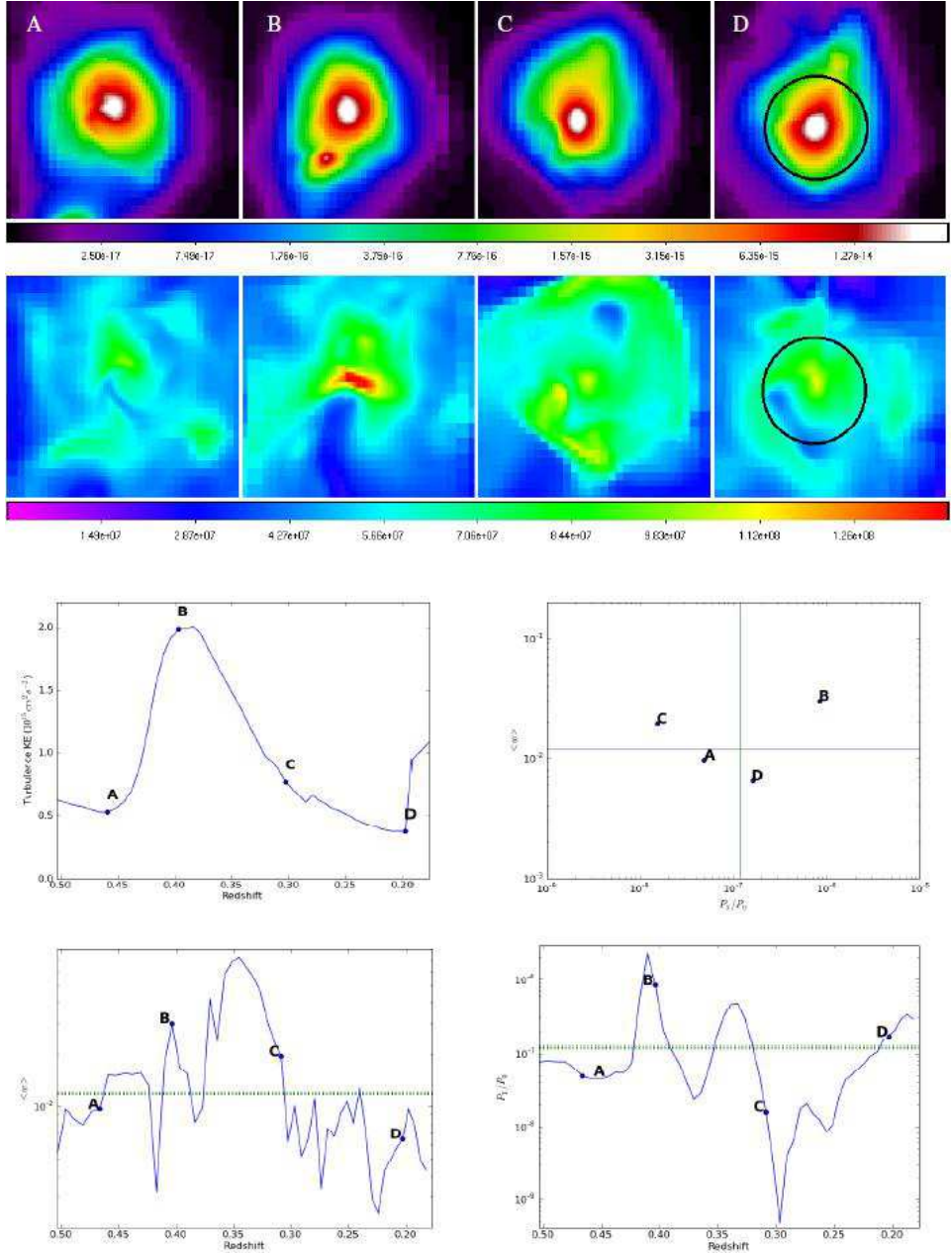}
\end{center}
\caption{Merger snapshots for a simulated galaxy cluster in our
  sample. Upper panels show images of the synthetic 0.3-8.0 keV X-ray surface
  brightness while the panels immediately below show the projected X-ray weighted
  temperature. Image panels are roughly 1 kpc across.  The black circle overlaid on the
final time step indicates the fiducial 500 kpc radius used in our
analysis for that timestep. The panels in the third row show the time evolution of
the turbulence kinetic energy (left) and the location of the cluster in the
$P_3/P_0-\langle w \rangle$ plane for the four time snapshots as indicated by
letter. Bottom panels show the time evolution of $P_3/P_0$ and
$\langle w \rangle$ with redshift; green dashed lines indicate the structure cuts used by \citet{cassano} to select radio halo clusters. The mass of the main cluster ranges from 4.0$\times 10^{14}
M_{\odot}$ at (A) to 5.0$\times 10^{14} M_{\odot}$ at (D). }
\label{halo4_merg}
\end{figure*}

\begin{figure*}
\begin{center}
\includegraphics[width=5.0in]{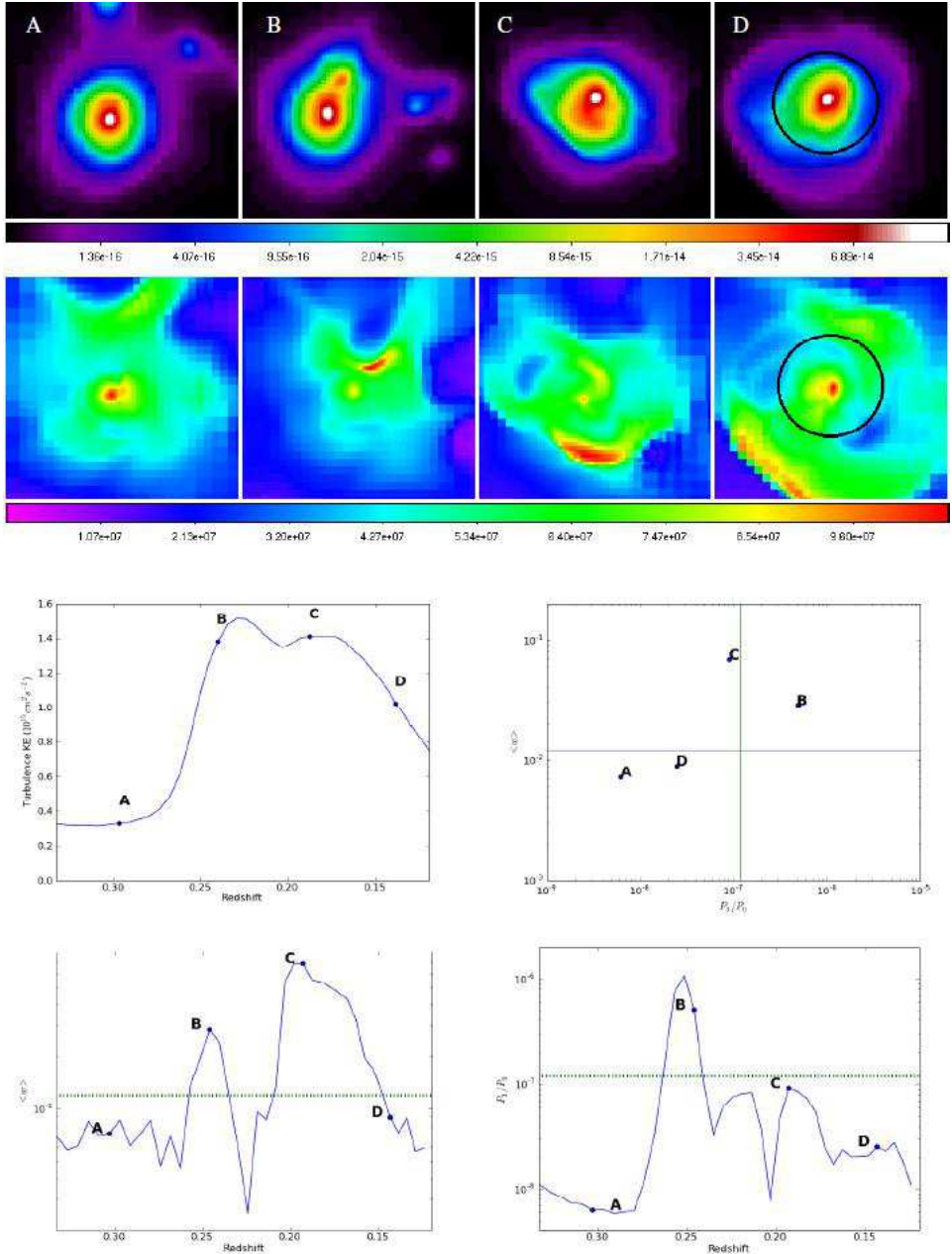}
\end{center}
\caption{Merger snapshots for a simulated galaxy cluster in our
  sample. Upper panels show images of the synthetic 0.3-8.0 keV X-ray surface
  brightness while the panels immediately below show the projected X-ray weighted
  temperature. Image panels are roughly 1 kpc across.  The black circle overlaid on the
final time step indicates the fiducial 500 kpc radius used in our
analysis for that timestep. The panels in the third row show the time evolution of
the turbulence kinetic energy (left) and the location of the cluster in the
$P_3/P_0-\langle w \rangle$ plane for the four time snapshots as indicated by
letter. Bottom panels show the time evolution of $P_3/P_0$ and
$\langle w \rangle$ with redshift; green dashed lines indicate the structure cuts used by \citet{cassano} to select radio halo clusters. The mass of the main cluster ranges from 2.8$\times 10^{14}
M_{\odot}$ at (A) to 3.5$\times 10^{14} M_{\odot}$ at (D).}
\label{halo14_merg}
\end{figure*}

\begin{figure*}
\begin{center}
\includegraphics[width=5.0in]{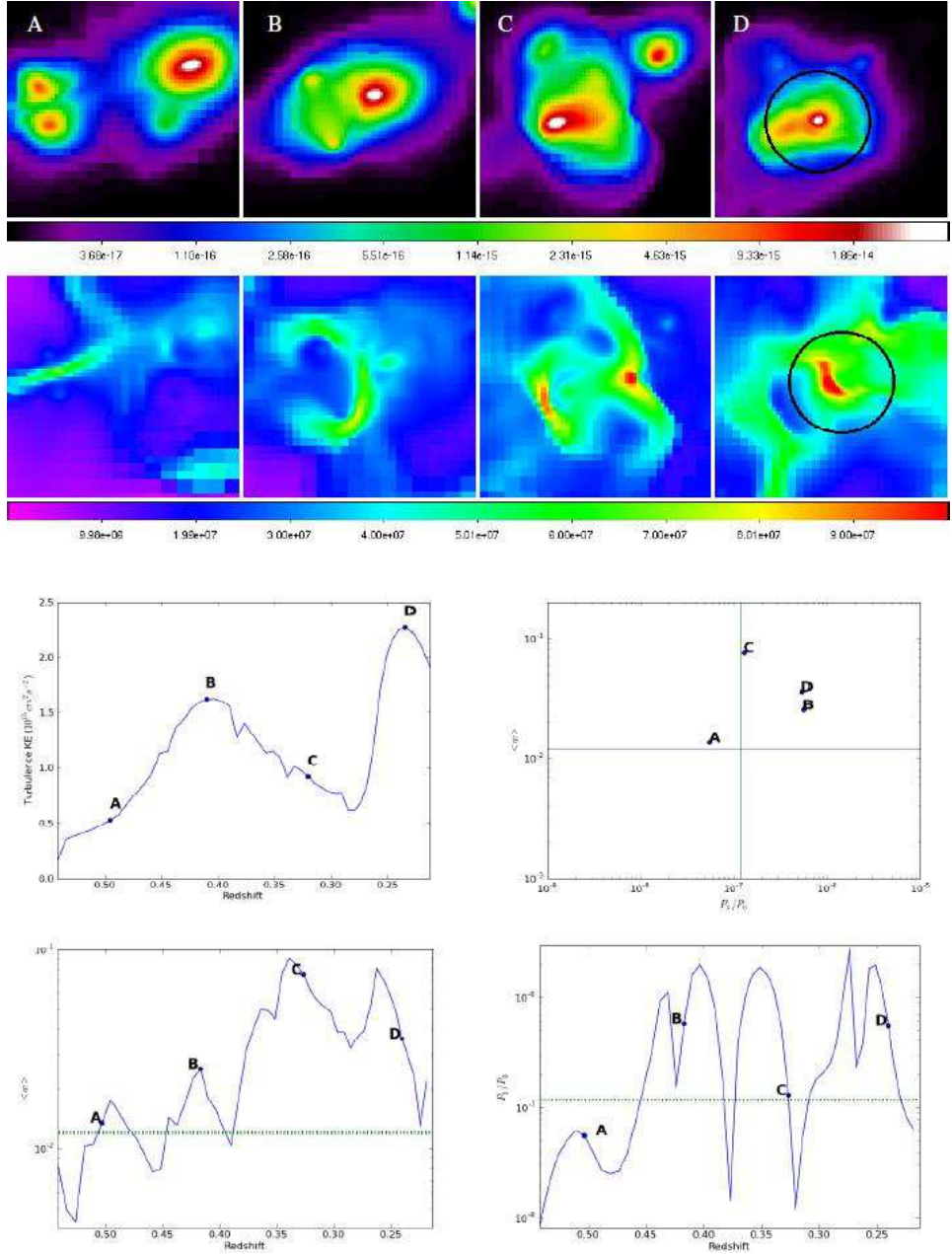}
\end{center}
\caption{Merger snapshots for a simulated galaxy cluster in our
  sample. Upper panels show images of the synthetic 0.3-8.0 keV X-ray surface
  brightness while the panels immediately below show the projected X-ray weighted
  temperature. Image panels are roughly 1 kpc across.  The black circle overlaid on the
final time step indicates the fiducial 500 kpc radius used in our
analysis for that timestep. The panels in the third row show the time evolution of
the turbulence kinetic energy (left) and the location of the cluster in the
$P_3/P_0-\langle w \rangle$ plane for the four time snapshots as indicated by
letter. Bottom panels show the time evolution of $P_3/P_0$ and
$\langle w \rangle$ with redshift; green dashed lines indicate the structure cuts used by \citet{cassano} to select radio halo clusters. The mass of the main cluster ranges from a very small 1.6$\times 10^{14}
M_{\odot}$ at (A) to more than double its original mass, 3.5$\times 10^{14} M_{\odot}$ at (D).}
\label{halo13_merg}
\end{figure*}

To further illuminate the effects of cluster mergers, in Figures \ref{halo4_merg}, \ref{halo14_merg}, and \ref{halo13_merg} we show three example
mergers from our simulated cluster sample which exhibit quite varied merger progessions. During the mergers, we have
measured the value of both $\langle w \rangle$ and $P_3/P_0$, as well
as the turbulence kinetic energy $k$.  In each figure, the images from
left to right in both the upper and second row are a time series of
X-ray surface brightness (upper) and X-ray weighted temperature
(second). The two plots in the second row from the bottom show the progression in time of the
TKE (left) and the position in the $P_3/P_0$-$\langle w \rangle$
plane. The bottom panels show the evolution of the individual
structure measures as a function of redshift. Letters indicate the position in the time series. In Figure
\ref{halo4_merg}, we see in panel A a subhalo approaching the main
cluster; in panel B, this subcluster is driving an obvious large shock
into the gas.  Point B in the TKE time series is associated with the
peak of the enhancement, and also the structure measures have moved
into the disturbed area of the $P_3/P_0$-$\langle w \rangle$ plane. As
the subcluster passes through the main cluster, the TKE declines
gradually, and the structure measures move back into the ``relaxed''
regions of the structure plot.  Between points B and C, the structure measures show the typical progression seen in cluster mergers: both $P_3/P_0$ and $\langle w \rangle$ show a peak near point B when the subcluster enters the 500 kpc aperture; as the cluster undergoes core passage between points B and C, the structure measures both dip; $P_3/P_0$ and $\langle w \rangle$ then show a second peak following core passage in the re-expansion phase before declining again.  Point C is near the end of re-expansion and outgoing shocks can be seen in the temperature map.  Compared to the structure measures, the TKE evolves smoothly throughout the merger.  At point D, the small remaining core of the subcluster can been seen re-entering the cluster, but this more minor event does not lead to as large a rise in the structure measures or the TKE.

Figure \ref{halo14_merg} shows a slightly more complicated
merger. First, we have two approaching subclusters (seen in the upper
center and upper right of panel A).  The first merger drives a large
shock, seen in the temperature image of panel B.  This again leads to
a spike in the TKE and a movement of the structure measures in to
the disturbed regime. Before the cluster can relax, the second,
apparently smaller subcluster plunges in, enhancing the TKE again and
again boosting the structure measures toward the disturbed regime (point C).  After an extended period, the TKE declines and the structure measures move in to the relaxed quadrant of the $P_3/P_0-\langle w \rangle$ plane (point D). 

Figure \ref{halo13_merg} shows a very complex scenario.  In panel A
we see two subhalos in the center left of the image which are in the process of merging. In panel B, the
subhalos have merged with each other and are in a re-expansion phase (note the merger shock in the
temperature image of panel A, and the outgoing shocks in B). They then pass by the main
cluster on either side, with the leading shock sweeping over the main
cluster.  Meanwhile, in panel C, note another approaching subcluster
to the upper right, which by panel D has plunged into the main
cluster.  This multiple merger, as seen in the context of the TKE and
structure measures, has an increase in $k$ during both merger events,
while the structure measures remain high throughout. This is a very
dynamic environment, where multiple mergers take place throughout the
cluster lifetime, leading to long periods of elevated TKE and
disturbed structure measures.

In general, we find that there is no "typical" cluster history.  Some clusters undergo complex mergers or progressions of successive mergers leading them to be disturbed over most of their history, while other clusters are nearly always relaxed, experiencing only one or two minor mergers from $z=0.9$ to the present.  This diversity highlights the importance of using cosmological scale simulations, since idealized mergers would not capture the full complexity of cluster evolution.

\subsection{Assessing the Turbulent Re-Acceleration Model}

It is clear from the cluster histories that the TKE can increase drastically during cluster mergers (as you would expect) and that clusters can display quite different TKE evolutionary histories.  In this section, we consider how the TKE correlates with the observed properties of radio halo clusters and if this is consistent with the turbulent re-acceleration model of radio halo production.  Specifically, we examine the TKE in clusters with and without X-ray structures consistent with those observed for radio halo clusters, and we look at the correlation between total TKE and cluster X-ray luminosity.

\subsubsection{Correlation of Cluster Structure and Turbulence}

\begin{figure*}
%\begin{center}
\includegraphics[width=6.5in]{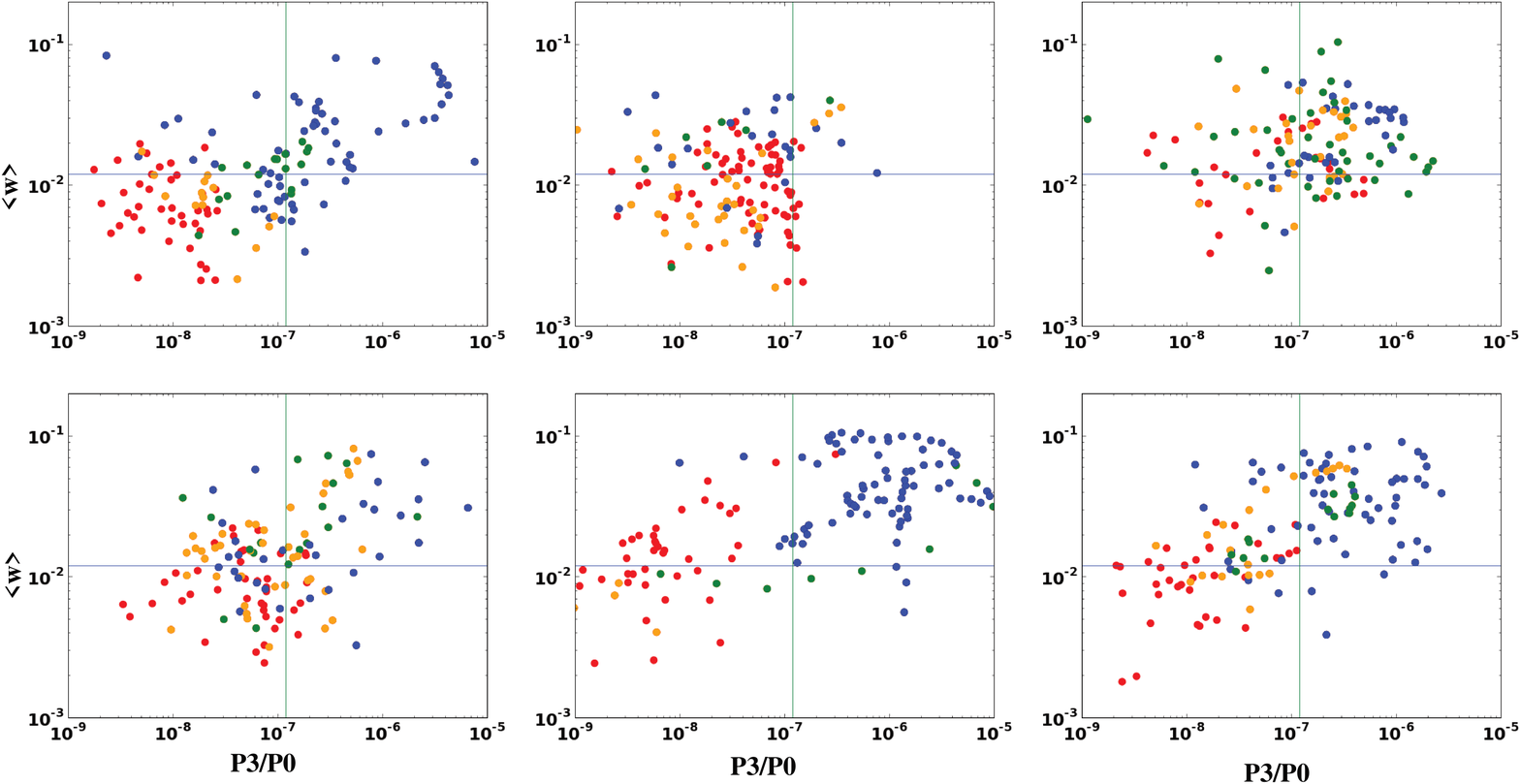}
%\end{center}
\caption{The $P_3/P_0$-$\langle w \rangle$ plane for projections of all time
  intervals where $M_{200} > 10^{14} M_{\odot}$ for the six
  $z$=0 clusters whose TKE time histories are shown in Figure
  \ref{all_ke_z}. In this plot, $\langle w \rangle$ is the
  centroid shift with the core excluded. Lines indicate the fiducial
  cut lines in $P_3/P_0$ and $\langle w \rangle$ in \citet{cassano}. Points are
  colored to indicate the relative level of turbulence kinetic energy
  for each time interval.  Red indicates intervals where $k <
  2k_{min}$, where $k_{min}$ is the minimum value of the TKE over the
  full time range for a given cluster. Yellow indicates $2k_{min} < k <
  3k_{min}$, green is $3k_{min} < k < 4k_{min}$, and blue is for $k >
  4k_{min}$.  }
\label{all_wp3}
\end{figure*}

One of the main relevant questions of our investigation is whether the
presence of enhanced TKE correlates with the value of the structure
measures for the X-ray images. We show
in Figure \ref{all_wp3} the $P_3/P_0$-$\langle w \rangle$ plane for
each of the clusters shown in Figure \ref{all_ke_z}. Each data point
in these images represents one time interval from the simulated data
for the cluster in question. The data points are colored by their
relative TKE, to indicate the level of turbulence at the time the
structures are measured. The colors go from red ($k <
  2k_{min}$) to yellow ($2k_{min} < k < 3k_{min}$), to green ($3k_{min} <
  k < 4k_{min}$), to blue ($k > 4k_{min}$).  In each case, $k_{min}$
  represents the minimum value of $k$ over the full time interval for
  which the cluster and its predecessors have $M > 10^{14}
  M_{\odot}$.  Using $k_{min}$ may lead to scatter in the results,
  however, it is not clear from any of the histories what the base
  value for $k$ should be.  In many objects, increases in $k$
  due to a new merger occur before $k$ has finished decreasing from prior
  mergers.  Certainly, by eye the visual effect of mergers lasts
  longer than the time between mergers for many of the objects. 
  
  The resulting correlations vary from cluster to cluster, but in
  general, Figure \ref{all_wp3} shows an obvious trend for more
  disturbed power ratios and centroid shifts to be
  accompanied by a higher TKE state. 
We can quantify this result further by examining the correlation of
cluster structure with the TKE in a global sense across all the
clusters. One key measurement is the amount of time spent in both an
elevated TKE state and in the upper right quadrant of the $P_3/P_0$ -
$\langle w \rangle$ plane.  Conversely, we can also check that the
lower left quadrant of this plot is more likely to be populated by
clusters with lower TKE. 

\begin{figure}
\begin{center}
\includegraphics[width=3.5in.]{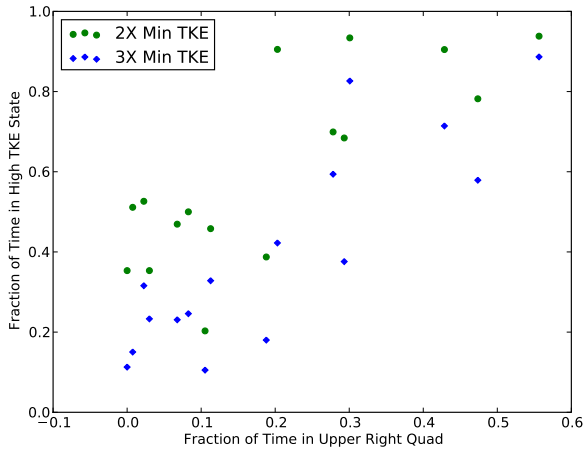}
\end{center}
\caption{Correlation of the time spent by our simulated clusters in elevated turbulence
  kinetic energy states with the fraction of time spent in the upper
  right (disturbed) quadrant of the $P_3/P_0$-$\langle w \rangle$
  plane (as defined by \citet{cassano}. Green points represent a value of $k > 2k_{min}$, blue
  indicates $k > 3k_{min}$. Note that some clusters spend almost the
  entire interval we studied in a high TKE state, while others spend
  almost no time at all in that state. }
\label{highke_upperright}
\end{figure}

\begin{figure}
\begin{center}
\includegraphics[width=3.5in]{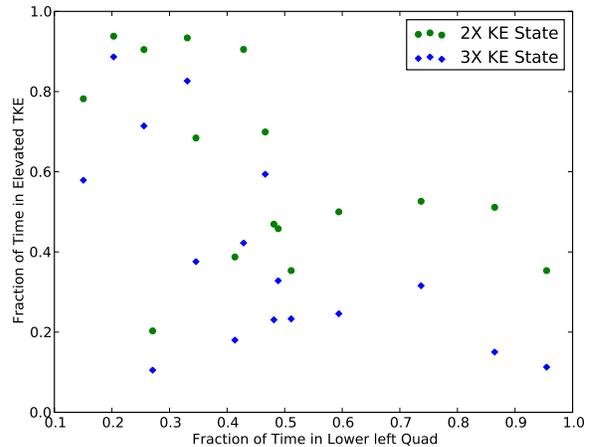}
\end{center}
\caption{Correlation of the time spent by our simulated clusters in elevated turbulence
  kinetic energy states with the amount of time spent in the lower
  left (relaxed) quadrant of the $P_3/P_0$-$\langle w \rangle$
  plane (as defined by \citet{cassano}. Green points represent a value of $k > 2k_{min}$, blue
  indicates $k > 3k_{min}$. }
\label{highke_lowerleft}
\end{figure}

Figures \ref{highke_upperright} and \ref{highke_lowerleft} show the
correlation between the turbulence kinetic energy and the location of
the cluster's projected structure measures.  These plots show the
fraction of time for each cluster that it spends in an elevated TKE
state (either twice or three times the minimum of $k$) versus the time
spent in the upper right quadrant or lower left quadrant of the
$P_3/P_0$-$\langle w \rangle$ plane. It is clear that these two
measures are correlated, though there is significant scatter.  To
first order, the clusters that spend more time in the elevated $k$
states spend more time with disturbed structure measures. It is also
evident that the converse is true, those that spend more time in
elevated $k$ states spend less time in the lower left of the structure
plane. It is also obvious from these Figures that the range of values
for fractions of time spent in various regimes are large.  Some
clusters spend a large fraction of time with elevated TKE and
disturbed morphology, while others spend almost no time in these
states. It is important to note that clusters do not follow some
standard evolutionary path, where they each spend a typical fraction
of time in elevated TKE states, but that each evolutionary track is
unique. There is a wide range of evolutionary histories, even in a
sample this small. 

Table \ref{tab:kvquads} shows, for the whole sample of clusters and
time outputs, the fraction of time that the structure measures lie in
both the relevant quadrants of the $P_3/P_0$-$\langle w \rangle$ plane
and in elevated TKE states.  The table shows clearly that clusters
that have disturbed structure measures are much more likely to have
elevated TKE.  While this correlation is not perfect, if a cluster has
disturbed structure measures, it has a 92\% chance of having its TKE
at least double the minimum over the interval in question.  Likewise, only a small fraction of clusters with structure measures indicating that they are relaxed have very elevated TKE.  Though
relaxed structure measures can not perfectly predict low TKE, a number
of factors contribute to this problem. First, the use of $k_{min}$ as
the base level for TKE almost certainly adds scatter to the
correlation. Also, as we have shown in earlier sections, there are
moments (particularly at merger core passage) where structure measures
look relaxed, yet the TKE is elevated. 

In \citet{cassano}, a strong case is
made that there is a real correlation between incidence of radio halos
and location in the planes described by combinations of the
concentration parameter, $\langle w \rangle$ and $P_3/P_0$.  The correlation found 
found here between cluster structure and elevated TKE lends credence to the claims of
  \citet{cassano} and others that turbulence could indeed be the
  energy source for the CRs that create the radio halo emission. 

\begin{table}
\caption{Fraction of Time with Elevated TKE for Location in Structure
  Plane}
\centering
\begin{tabular}{lccc}
\hline
\hline
$P_3/P_0$-$\langle w \rangle$ Quad &  2x  TKE & 3x TKE & 4x TKE \\
\hline
Upper Right & 0.92 & 0.77 & 0.58 \\
Lower Left & 0.44 & 0.19 & 0.13 \\
\hline
\end{tabular}
\label{tab:kvquads}
\end{table}


\subsubsection{X-ray Luminosity versus Turbulence Kinetic Energy}
\begin{figure}
\includegraphics[width=3.5in]{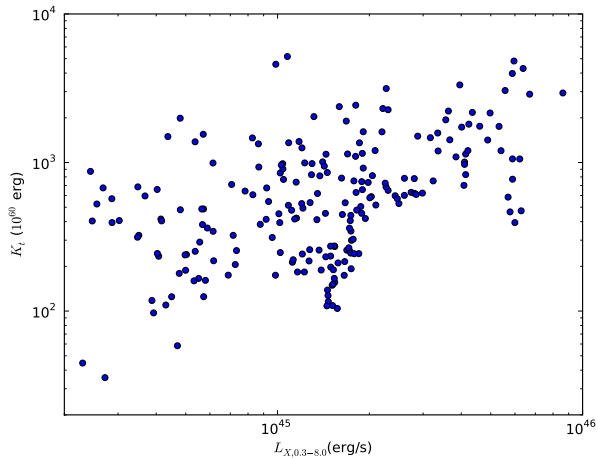}
\caption{Synthetic 0.3-8.0 keV X-ray luminosity versus the total turbulence kinetic
  energy ($M_{gas}k$) for a randomly selected 10\% of the time outputs for all clusters and their
progenitors in our sample. $L_x$ is determined from the projected mock
images, $K_t$ is extracted from the three-dimensional simulation
data. }
\label{Lx-k}
\end{figure}

Under the assumption that turbulent bulk motions provide the energy
source for the CR electrons radiating as radio halos, we plot the
total (not specific) turbulence kinetic energy $K_t = M_{gas}k$ against the
synthetic X-ray luminosity of our simulated clusters in Figure \ref{Lx-k}.  The X-ray
luminosity is extracted from the projected maps used for the structure
measures, and the turbulence kinetic energy is integrated in each
cluster volume (radius of 500kpc).  The relation is shown in Figure \ref{Lx-k} for a randomly selected 10\% of the
timesteps in each cluster history (to avoid the appearance of
evolutionary tracks). We can not expect identical behavior to the
$L_x-P_{1.4 GHz}$ relation for observed clusters due to the complicating
factors of the magnetic field strength and distribution, and the aging
of the accelerated particles. However, it is interesting to check
whether the ``switch'' for on/off states of radio halos results
directly from a quick transition in turbulence state. Previous work
\citep[e.g.,][]{paul} suggests that turbulent pressure is long-lived
in the ICM after mergers. Our relation shows no obvious evidence for
bimodality in the relation between X-ray luminosity and turbulence
kinetic energy. Indeed, our result suggests that the
transition from turbulent to relaxed is gradual, as indicated by our
earlier time histories. Our analysis shows that there is an order of
magnitude spread in the value of $K_t$ for a given X-ray
luminosity. This result is consistent with \citet{rudnick}, which
indicates that the value of $P_{1.4}$ can take on a range of values
for a given $L_x$, under some upper envelope.  %If there is a bimodal $L_x - P_{1.4}$ for radio halos, in the turbulence re-acceleration model it must be accounted for by rapid synchrotron cooling ($\tau \sim 100$ Myr). 
 It is possible that once a large population of radio halos can be detected at longer wavelengths (given their steep radio spectra) any clear bimodality in the
$L_x$-$P_R$ relations will be washed out. Also, though the $L_x - K_t$
relation presented here is not bimodal, there are other relevant physical
effects on the radio emission (magnetic field strength and variation, presence and spectrum of a pre-existing electron population) which are not included here that could induce strong effects on the resulting
$L_x - P_R$ ratio. For example, \citet{brunetti09} suggest that as the
turbulent kinetic energy decays following a merger, the the
suppression of the synchrotron emissivity in a fixed radio band will
be strong.  This also suggests that a factor of two difference in the
value of $k$ will lead to a factor of 10 suppression in the
synchrotron luminosity at 1.4 GHz. In that case, a continuous
distribution in $K_t$ for a given $L_x$, as appears in our figure,
will still show a sharp boundary, leading to the observed bimodality.

\section{DISTRIBUTION AND DURATION OF TURBULENCE IN SIMULATED CLUSTERS}

If in fact turbulent re-acceleration is responsible for generating the relativistic electrons powering radio halos, then the evolution of the TKE in our simulated clusters can tell us about the possible lifetime of radio halos and the frequency with which clusters are expected to host radio halos.  In this section, we examine the typical duration of episodes of elevated turbulence (i.e. the merger relaxation time) and the fraction of clusters at a given time that are in elevated TKE states.

\subsection{Lifetime of Merging Events}

\begin{figure}
\begin{center}
\includegraphics[width=3.5in]{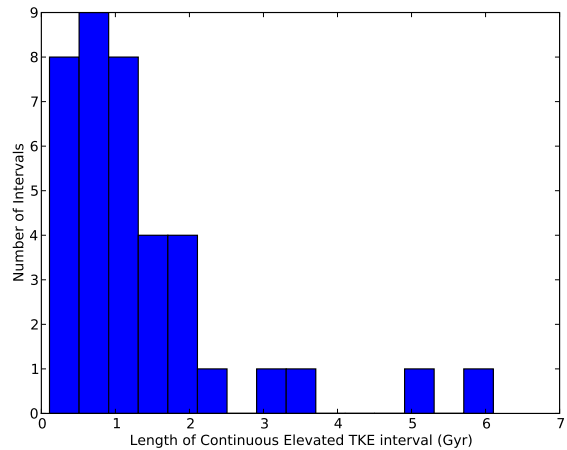}
\end{center}
\caption{Histogram of all continuous time intervals for clusters in
  our sample where the turbulence kinetic energy inside r=500 $kpc$ is elevated to $k >
  2k_{min}$. The median interval length is 1.12 $Gyr$. }
\label{cont_2x}
\end{figure}

We have run an analysis to determine the typical timescale for
clusters in our sample to be in elevated TKE states. We take every
cluster's time history and identify all time intervals over which the
TKE is continuously elevated to at least 2$k_{min}$. The histogram of
those intervals is shown in Figure \ref{cont_2x}.  The median length
of time that a cluster spends in a continuously elevated TKE state by
this definition is $t_{ele} = 1.13$ Gyr. If we redefine the intervals
by elevation to 3$k_{min}$ or above, then $t_{ele} = 0.77$ Gyr. For $k
> 4k_{min}$, $t_{ele} = 0.77$ Gyr. While it may not be obvious why the
median interval for the highest two $k$ bins should be the same, the
rough shape of the distribution is preserved at these ranges, though
the normalization has declined. So it is safe to say from our
sample that the median time interval over which the TKE is elevated
inside r =500 kpc is around 1 Gyr, or slightly less depending on the
threshold used.  However, Figure \ref{cont_2x} also shows a tail to very long intervals of elevated TKE with some clusters having time periods of 5 or more Gyrs with continuously elevated TKE.  This result again shows the diversity of cluster histories: in the turbulent re-acceleration model some clusters would be expected to host radio halos over most of their lifetime while other clusters never would.

Continuous time intervals for elevated structure measures show a
different result. The median value for intervals of $P_3/P_0$ elevated
above the \citet{cassano} limit is much shorter than the elevated $k$
intervals at $t_{ele} = 0.33 Gyr$. For the centroid shift we see a
similar result of $t_{ele} = 0.39 Gyr$.  This result is not
particularly surprising given the time histories shown in Figures
\ref{halo4_merg}, \ref{halo14_merg} and \ref{halo13_merg}. The values of the
structure measures are more highly variable over short time scales
than the value of $k$, and can show strong variations at core passage
for instance.  If we identify by eye time periods with low structure measures associated to core passage events and interpolate over them, a longer typical timescale for elevated structure measure of around $0.5 Gyr$ is found, though this procedure is subjective.

\subsection{Fraction of Clusters with Elevated Turbulence Kinetic
  Energy}
\begin{figure}
\includegraphics[width=3.5in]{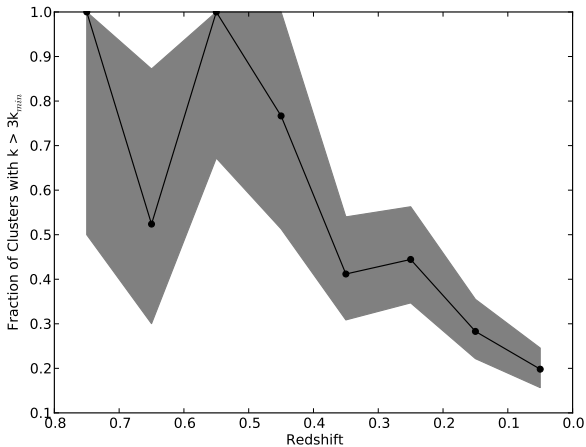}
\caption{The fraction of simulated massive ($M > 5 \times 10^{14}
  M_{\odot}$) clusters as $f(z)$ that have $k >
  3k_{min}$. %If turbulent re-acceleration is the method of generating radio halos, this result may track with the fraction of clusters with giant RHs (at least at low redshift where IC losses are smallest). 
  Shaded region shows Poisson error bars due to the
  finite sample of simulated clusters at each redshift. }
\label{k_vs_z}
\end{figure}

If the energy source for giant radio halos is turbulent
re-acceleration, an interesting measure is the fraction of objects
with elevated turbulence kinetic energy. In Figure \ref{k_vs_z}, we
show the fraction of massive clusters (M$\geq 5\times 10^{14}M_{\odot})$ as a function of redshift that have $k$
elevated by an integer factor above the minimum over their
history. The fraction is a noisy, but gradually declining function of
time depending on what threshold of $k$ is used. For $k > 3k_{min}$ as shown in 
Figure \ref{k_vs_z}, the fraction of clusters with elevated TKE at
high redshift ($z > 0.4$) is noisy, but high (50-100\%), and declines
to roughly 20\% at $z=0$.  The trend for values of $k > 2k_{min}$
and $k > 4k_{min}$ is similar, but the normalization changes. For $k >
2k_{min}$ the typical fraction of clusters declines from 60-100\% at
the higher redshifts to 35-50\% at $z$=0. For $k > 4k_{min}$ the
fraction is 50-100\% at high $z$, with a drop to around 8\% at $z$=0. Our results
imply that the fraction of massive clusters that are energetically
favorable for producing CRs through turbulent re-acceleration is
declining with redshift.

Observational searches
find a fraction of roughly 30\% of clusters at moderate redshifts ($0.2<z<0.4$) hosting radio
halos \citep{venturi}. This result will obviously depend on the survey depth and should also depend strongly on the frequency of
observation, as steep spectrum radio sources are much brighter at
longer wavelengths. In addition, we should point out that radiative
losses to relativistic particles due to inverse Compton (IC)
interactions with the CMB scale as (1+$z)^4$, strongly limiting the
possibility of radio halos at the higher redshifts \citep{cassano05,cassano06}. 
Given the uncertainty in the energetic
threshold for TKE to produce a radio halo, our range for the fraction
of clusters with elevated TKE in the redshift range of most observed
radio halos ($0.2 < z < 0.4$) of 20-50\% is broadly consistent with
the observed ratio. 

\section{DISCUSSION}
Observations show a connection between the detection of diffuse, large-scale radio halo emission and the presence of merging activity in clusters \citep[e.g.][]{buote, cassano}.  In particular, these studies show a correlation between observed X-ray cluster structure, quantified based on the centroid shift and the power ratios, and the presence of radio halo emission.  While the origin of radio halos is still a matter of debate, it has been suggested that merger generated turbulence could accelerate a pre-existing relativistic electron population (from supernovae, AGN, previous merger/shock activity) to energies sufficient to produce the radio emission.  

In this paper, we study the evolution with time of massive clusters formed within a numerical hydrodynamic, cosmological simulation and investigate whether a subset of the X-ray and radio observational evidence from clusters is consistent with the turbulent re-acceleration model for radio halos.  In particular, we track the evolution of random bulk ICM motions, the kinetic energy of which is dissipated through shocks and the turbulent cascade.  We study the relationship between turbulence kinetic energy and X-ray structure measures, how these evolve during mergers, and whether the level of turbulence in simulated clusters with radio halo-like cluster structures supports the turbulent re-acceleration model.  

We show that the distribution of the X-ray structure measures (centroid shift $w$ and third order power ratio $P_3/P_0$) used
by \citet{cassano} to separate radio halo and non-radio halo clusters are reproduced fairly well in our simulated cluster
sample and that a similar fraction of simulated clusters have "disturbed" X-ray structures compared to observations.   
We also find that clusters in the disturbed regimes of the structure plane are highly likely to
have elevated turbulence kinetic energy (at the scales we
resolve). For some
clusters, we have shown this graphically (Figure \ref{all_wp3}).
Quantitatively, clusters with structure measures in the ``disturbed''
part of the structure plane have a 92\% chance of having an elevated
$k$ at least two times above the minimum over the cluster lifetime.  Though
individually, there is a significant amount of variation in
the structure measures for a given value of $k$, as a group they
correlate strongly. \citet{cassano} suggest that the fact that radio halo clusters are more disturbed in terms of their structure measures results from a difference in the amount of turbulence in disturbed versus relaxed clusters and that turbulent re-acceleration of CR electrons is the source of radio halos in galaxy clusters. The strong correlation between cluster structure and elevated turbulence in simulated clusters supports this theory.

The value of the turbulence kinetic energy $k$ undergoes large
variations during mergers of factors of a few to several. Mergers are
typically well-defined in $k(z)$, with $k$ smoothly increasing and then decreasing through the merger, though additional mergers can (not infrequently) occur before the system can fully relax. The
time evolution of the structure measures during mergers are not as well defined and are a strong function of merger phase. Both $P_3/P_0$ and $\langle w
\rangle$ are noisier than $k$ as a function of time in our simulated
clusters. A classic example in many of our simulated clusters is the
sharp reduction in the values of the structure measures as a merger proceeds
through the core passage of a subcluster. Though the turbulence kinetic
energy is high at this stage, the object could be called ``relaxed''
using the structure measures and significant substructure is not visible in the X-ray image. The result of this difference is that
the time scales for continuously elevated $k$ are of order 2-3 times
longer than those for elevated structure measures. The median interval
for enhanced $k$ to twice the minimum for a given cluster is
approximately $\tau$=1 Gyr. For $P_3/P_0$ and $\langle w \rangle$,
this number is closer to $\tau$= 0.33 Gyr. For these reasons, translating from
the observed structure measures to the dynamical state, or more
specifically to the energetics of the bulk motion of the gas, has significant scatter for individual clusters. However, statistically these properties correlate well.

Though prior studies suggest turbulent kinetic energy is
very long-lived in clusters \citep[e.g.,][]{paul}, we show that $k$
drops significantly in cluster volumes over much shorter timescales
($\sim 0.5$ Gyr).  However, these results may be consistent, given
that the \citet{paul} result uses a different criterion (turbulent
pressure $>5$\% of the total pressure) than we do for elevated
turbulence. The relatively quick drop in TKE, coupled to short synchrotron cooling
times, may account for the sharp difference in X-ray structure for RH
versus non-RH clusters.

An interesting result of this work is that we find that many clusters
spend a large fraction ($\sim$80+\%) of their time in disturbed
states, while others spend very little time ($\sim$10\%) in such states.
Thus, it may be the case that some galaxy clusters are rarely, if ever
radio halo clusters during their lifetime, while others almost
always host radio halos.  While the typical timescale for the turbulence kinetic energy to be elevated is $\sim 1$ Gyr, these periods can be as long as 5 or more Gyrs in cluster which experience complex or repeated mergers.

Finally, we find no obvious evidence for bimodality of the $L_x$-$K$
relation (where $K$ could be a proxy for radio power) in our simulated
sample of clusters.  However, this is not necessarily surprising given
that there are a number of reasons why such a gap may exist in the observed
 $L_x$-$P_{1.4}$ relation that are unrelated to the dynamical state of
 the gas.  For example, we have not modeled magnetic field variations, and the turbulent re-acceleration
 mechanism requires a pre-existing population of CRs on which to
 operate, both of which may vary from cluster to cluster.
 \citet{brunetti09} have shown that the relationship between the
 kinetic energy in turbulence and the fixed band radio luminosity is
 not linear.  Their work suggests that a relatively small drop in TKE
 results in a large change in the synchrotron luminosity, which could
 easily account for the observed bimodality, yet still be consistent
 with our result here.
 To fully understand this issue, it is
 necessary to follow the magnetic fields and CR particles explicitly
 in the simulation to see the true evolution of the radio emission.  A full treatment of simulating a large sample of
galaxy clusters and their radio emission must handle the injection of
CRs into the ICM from reasonable sources like AGN, SNe, and large-scale accretion shocks.  This work we defer to
upcoming simulations using a new MHD+CR code.
\section*{Acknowledgments}
EJH acknowledges the support of Fermi Guest Investigator grant 21077.
EJH acknowledges very helpful discussions with Rossella Cassano, Simona Giacintucci, Marcus
Brueggen, Brian O'Shea, Sam Skillman and Uri Keshet. TEJ acknowledges
support from NASA grant NNX09AT83G. Computations described in this work were performed using
the \texttt{Enzo} code developed by the Laboratory for Computational
Astrophysics at the University of California
in San Diego (http://lca.ucsd.edu) and by a community
of independent developers from numerous other institutions.
The \texttt{YT} analysis toolkit was developed primarily
by Matthew Turk with contributions from many other
developers, to whom we are very grateful. Computing time was provided by
TRAC allocation TG-AST100004. The authors acknowledge the
Texas Advanced Computing Center (TACC) at The University
of Texas at Austin for providing HPC resources
that have contributed to the research results reported
within this paper. URL: http://www.tacc.utexas.edu. EJH acknowledges
the wifi connections provided by the Massachusetts Bay Transportation
Authority's Commuter Rail (Franklin Line) in enabling significant
portions of this work to be done in transit. 
\bibliographystyle{apj}
%\bibliography{ms,extra}

\end{document}